\documentclass[twocolumn,showpacs,preprintnumbers,amsmath,amssymb]{revtex4-1}

\usepackage{xspace}
\usepackage{epsfig}
\usepackage{amsfonts}
\usepackage{graphicx}
\usepackage{url}
\usepackage[dvipdfm,%
              colorlinks,%
          linkcolor=blue,%
        citecolor = blue,%
              hyperindex,%
        plainpages=false,%
         urlcolor = blue,%
       pdfstartview=FitH]{hyperref}
\usepackage{url}

\begin{document}

%\preprint{AIP/123-QED}

\title{Stark-Effect Scattering in Rough Quantum Wells}

\author{Raj K. Jana}
\email[]{rjana1@nd.edu}
%\affiliation{ Department of Electrical Engineering, University of Notre Dame, Indiana 46556, USA.}
\author{ Debdeep Jena}
\affiliation{ Department of Electrical Engineering, University of Notre Dame, Indiana 46556, USA.}

\date{\today}% It is always \today, today,
             %  but any date may be explicitly specified

\begin{abstract}

A scattering mechanism stemming from the Stark-shift of energy levels by electric fields in semiconductor quantum wells is identified.  This scattering mechanism feeds off interface roughness and electric fields, and modifies the well known `sixth-power' law of electron mobility degradation.  This work first treats Stark-effect scattering in rough quantum wells as a perturbation for small electric fields, and then directly absorbs it into the Hamiltonian for large fields.  The major result is the existence of a {\em window} of quantum well widths for which the combined roughness scattering is minimum.  Carrier scattering and mobility degradation in wide quantum wells are thus expected to be equally severe as in narrow wells due to Stark-effect scattering in electric fields.

\end{abstract}

\pacs{Valid PACS appear here}% PACS, the Physics and Astronomy
                             % Classification Scheme.
%\keywords{Suggested keywords}%Use showkeys class option if keyword
                              %display desired
\maketitle

High-mobility 2DEGs have proven invaluable for fundamental discoveries in condensed matter physics such as the quantum Hall effect, quantized conductance, and ballistic transport among many others \cite{AndoRMP82, BookQHE} and thus continuous improvement in the mobilities and mean free paths of carriers are highly desirable.  For high-speed and low-power field-effect transistors (FETs), a high degree of vertical scaling is essential to support lateral (gate length) scaling, requiring one to move towards highly confined 2DEGs in ultrathin quantum wells such as in Silicon-on-Insulator (SOI) and III-V Quantum Well (QW) FETs \cite{EsseniITED01,SunIEDL09}) to avoid short-channel effects.  Thus, interface roughness scattering assumes increasing importance in high-performance transistors.  In their seminal work in 1987 Sakaki {\em et al.} identified the importance of interface roughness scattering on electron transport in 2-dimensional electron gases (2DEGs) confined in narrow quantum wells \cite{SakakiAPL87}.  They showed that in the presence of quantum well width ($L_{w}$) variations in the 2D plane, the electron mobility limited by interface roughness (IR) scattering degrades in thin wells as the {\em sixth power} of the well-width ($\mu_{IR} \sim L_{w}^{6}$).

Sakaki {\em et al.} assumed a QW with no electric field \cite{SakakiAPL87}.  In typical QW FETs, the electric field indeed goes to zero when the carriers are depleted (when the device is in the `off' state), and increases to high values in the `on' state of the device.  For high-performance devices, a high 2DEG density is essential for boosting the drive current - which results in high electric fields in the QW.  In this work, we show that the electric field in the QW leads to an enhanced quantum-confined `Stark-effect' scattering that feeds off interface roughness, and degrades electron mobility in rough quantum wells.  We first evaluate the effect of Stark-effect scattering in a QW in cases where the potential fluctuation due to the electric field is small enough to be treated as a perturbation.  Then, we discuss situations where the field is so large that a perturbative treatment does not do justice, and a modified treatment that treats IR+Stark-effect scattering on equal footing captures the role of this mobility degradation mechanism.  We note that this form of scattering is incorporated in recent numerical approaches (see \cite{JinITED07}), therefore the purpose of this work is to offer an analytical framework for clear visualization of the physics and for ease of design.

\begin{figure}[b]
\includegraphics*[width=87 mm]{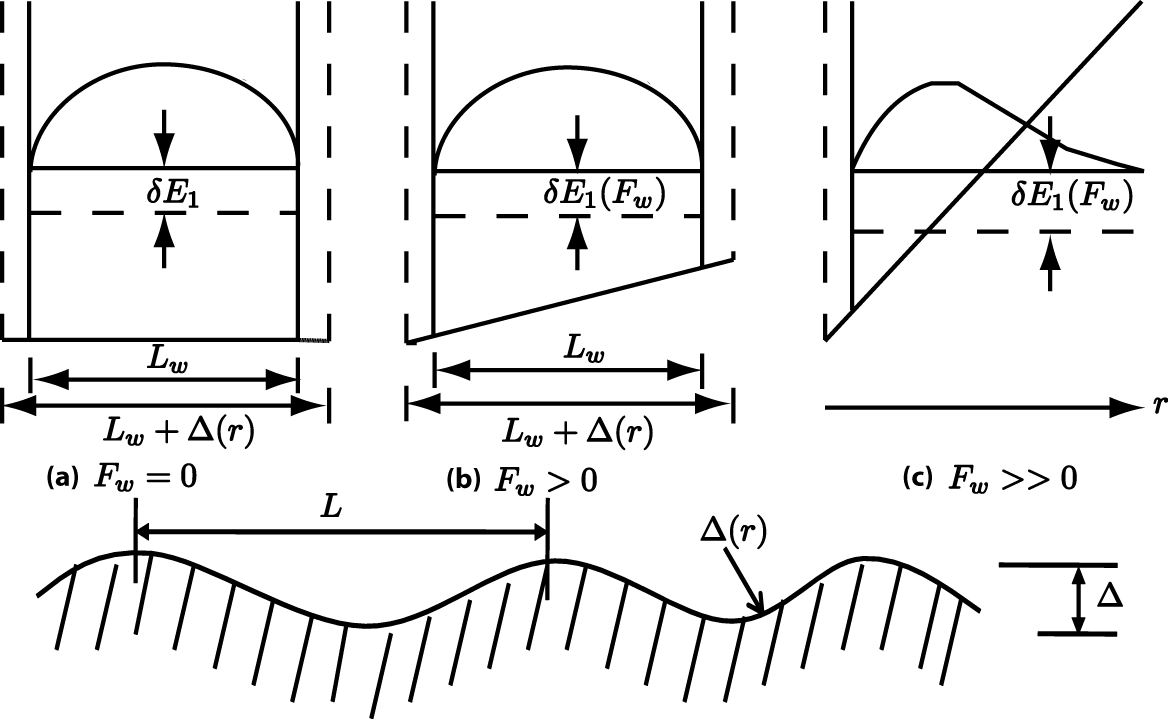}
\caption{%
Schematic figure illustrating interface roughness.  (a) and (b) are square QWs without and with an electric field respectively, and (c) is the case of a triangular QW at a high field.  Dashed lines indicate wider wells and corresponding eigenvalue fluctuations.  Interface roughness parameters $\Delta$ and $L$ are illustrated.}
 \label{Fig1}
\end{figure}

The central problem is illustrated in Fig.\ref{Fig1}.  Following \cite{SakakiAPL87}, the QW is visualized to be of width $L_{w}(r) = L_{w} + \Delta(r)$, where $r=(x,y)$ is the in-plane coordinate, $\Delta(r)$ is the fluctuation function with a correlation $\langle \Delta(r)\Delta(r+r') \rangle_{r} = \Delta^{2}\exp[-(r'/L)^{2}]$ and mean $\langle \Delta(r) \rangle_{r} = 0$.  The QW width fluctuation is parametrized by the height $\Delta$ and the in-plane correlation length $L$ as shown in Fig.\ref{Fig1}.  Assuming an infinite quantum well, the ground-state ($n=1$) energy at {\em zero vertical field} ($F_{w}=0$) is $E_{1}(F_{w} = 0) = \pi^2 \hbar^2 /2m^{\star} L_{w}^2$, where $\hbar = h/2\pi$ is the reduced Planck's constant, and $m^{\star}$ is the electron effective mass.  Variations in the QW width by $\Delta(r)$ changes the ground state energy by
\begin{equation}
\delta E_{1}(r, F_{w} = 0)  =  \frac{\partial E_{1} (0)}{\partial L_w} \Delta(r)= -  \frac{\hbar^2\pi^2}{m^*L_w^3}\Delta(r),
\end{equation}
which was the premise of \cite{SakakiAPL87}, leading to a $\sim L_{w}^{6}$ mobility variation.  In the presence of an electric field in the well, the ground state energy shifts.  This quantum-confined Stark-effect shift manifests in spectral shifts in optical transitions, but has not yet been related to transport properties.  The energy shift of the ground ($n=1$) state is obtained using 2nd-order perturbation by summing the contributions due to interactions with states $n=2,3, ...$, and is given by $ E_{1} (F_{w}) = E_{1}(F_{w}=0) - 24(2/3\pi)^{6} e^{2} m^{\star} L_{w}^{4}F_{w}^{2}/\hbar^{2}$ \cite{BookFox}, where $e$ is the electron charge.  Including the Stark-shift results in an increased scattering potential
\begin{equation}
\delta E_{1}(r, F_{w})  = - [\frac{\hbar^2\pi^2}{m^{\star} L_w^3} + 96(\frac{2}{3\pi})^{6} \frac{e^{2} m^{\star} L_{w}^{3} F_{w}^{2}}{\hbar^{2}}] \Delta(r),
\label{scatpotential}
\end{equation}
where the dependence on the electric field appears explicitly.  The scattering potential thus takes the form $W(r) = \delta E_{1}(r, F_{w}) = \mathcal{F} \Delta(r)$, where $\mathcal{F} = A/(m^{\star} L_{w}^{3}) + B \cdot m^{\star} L_{w}^{3} F_{w}^{2} $ is an effective force and $A, B$ are constants.  In the presence of an electric field, the Stark-effect scattering potential {\em increases} with QW width as $L_{w}^{3}$, which can be understood from Fig \ref{Fig1}(b) since there is a larger potential drop due to roughness.  The net scattering potential in Eq.\ref{scatpotential} thus goes through a {\em minimum} for a critical $L_{w}^{0}(F_{w})  \sim 14/ (m^{\star} F_{w})^{1/3}$ nm, where $m^{\star}$ and $F_{w}$ are normalized to the free electron mass $m_{0}$ and 10 kV/cm respectively for numerical evaluation.  The mobility limited by combined IR + Stark-effect scattering is expected to be a maximum for this QW thickness.  We further note that since the zero-field `Sakaki' term depends on effective mass as $\sim 1/m^{\star}$ and the `Stark' term as $\sim m^{\star}$, the Stark term will dominate for hole gases.

\begin{figure*}[t]
\begin{center}
\leavevmode
\epsfxsize=7in
\epsffile{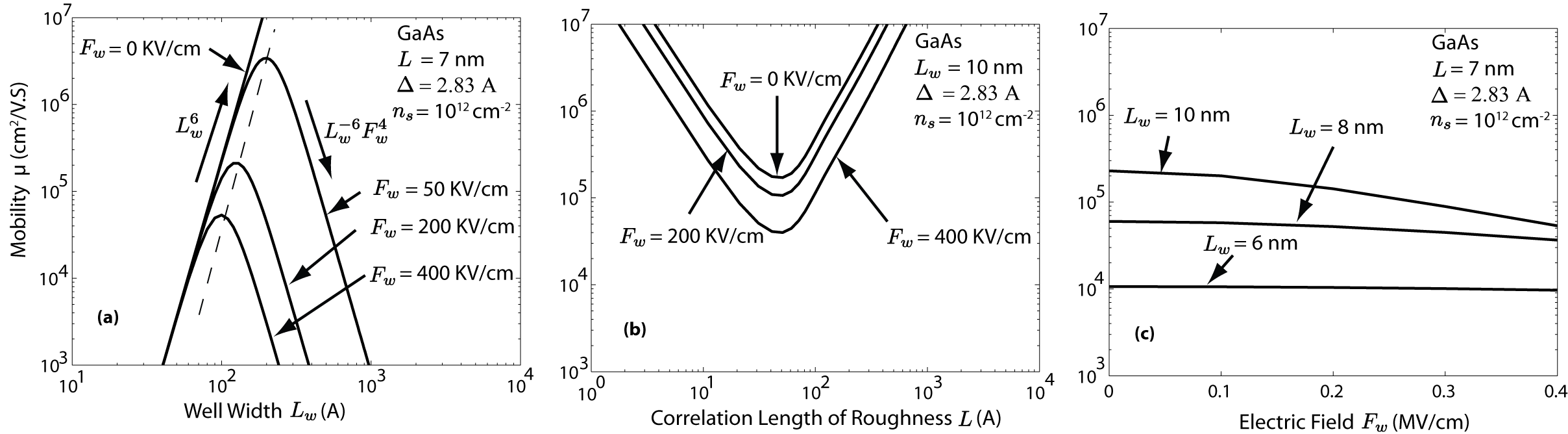}
\caption{(a) Mobility as a function of well width $L_w$ for various electric fields $F_{w}$.  Dashed line indicates a decreasing peak mobility with increasing applied electric field. (b) Mobility with correlation length $L$ for various electric fields. (c) Mobility vs. electric field for various QW thicknesses. }
\label{Fig2}
\end{center}
\end{figure*}

Electron mobility is calculated using Fermi's golden rule in the Born-approximation, which requires the squared 2D Fourier-transform of the scattering potential $W(r)$.  For scattering from state $| {\bf k} \rangle \rightarrow | {\bf k'} \rangle$, it evaluates to $|W(q)|^{2} = \pi \mathcal{F}^{2} \Delta^{2} L^{2} \exp[- (qL)^{2}/4] $, with $q = |{\bf k - k'}| = 2|{\bf k}| \sin( \theta/2)$, where $\theta$ is the angle between the 2D wavevectors ${\bf k}$ and ${\bf k'}$.  The momentum relaxation rate in the relaxation-time approximation of the Bolzmann transport equation is then
\begin{equation}
\frac{1}{\tau_m (k)}= \frac{2\pi}{\hbar} \int \frac{d^{2} {\bf k'}} {(2\pi)^2}  |\frac{W(q)}{\epsilon_{2D} (q)}|^{2} (1-\cos\theta) \delta(E_{k'} -E_{k}),
\end{equation}
where $\epsilon_{2D} (q)=1+ q_{TF}/q$ is the 2D screening function, $q_{TF}= m^{\star} e^2/ 2 \pi \hbar^2 \epsilon_s $ is the Thomas-Fermi wave vector, and $\epsilon_{s} = \epsilon_{0} \epsilon_{r}$ is the dielectric constant of the semiconductor \cite{BookDavis}.  Moving to radial coordinates and using the property of the delta function, the integral converts to one over scattering angles -
\begin{equation}
\frac{1}{\tau_m (k)}= \frac{m^{\star} \mathcal{F}^2 \Delta^2 L^2 }{2 \hbar^3} \int_{0}^{2\pi} d\theta \frac{e^{-k^2 L^2  sin^2\frac{\theta}{2}}}{\epsilon_{2D}^{2}(2k\sin\frac{\theta}{2})} (1-\cos\theta).
\label{sqwellresult}
\end{equation}

For typically degenerate 2DEG carriers, transport occurs at the Fermi level, and averaging the momentum relaxation rate over the carrier distribution amounts to evaluating it at the Fermi wavevector determined by the 2DEG carrier density $|{\bf k}| = k_{F} = \sqrt{2 \pi n_{s}}$.  The net electron mobility is then obtained as $\mu_{IR} = e  \tau_{m}(k_{F})/m^{\star}$.  Using the same example as in \cite{SakakiAPL87}, we choose 2DEGs in GaAs QWs (with $m^{\star} = 0.067 m_{0}$, $\Delta= 2.83 A$,  $\epsilon_{r}=12.9$) to illustrate the effect of Stark-effect scattering.  The results are shown in Fig \ref{Fig2}.

Fig \ref{Fig2}(a) shows the net interface roughness scattering rate as a function of the QW width for various strengths of the electric field in the QW.  For $F_{w}=0$, the result essentially is the same as Sakaki {\em et al's} result \cite{SakakiAPL87}, showing a monotonic increase as $\mu \sim L_{w}^{6}$, and Stark-effect scattering is absent.  Turning on $F_{w}$ causes the mobility to peak at $L_{w}^{0}(F_{w})$, and then drop with increasing $L_{w}$ as $\mu \sim L_{w}^{-6}$.  In this regime, the Stark-effect scattering dominates.  This behavior is explained by the two competing terms in Eq. \ref{scatpotential} as a function of $L_{w}$.  The QW width at which the maximum mobility is reached decreases with increasing $F_{w}$.  Fig \ref{Fig2}(b) shows the effect of the correlation-length of fluctuations at various values of $F_{w}$.  The mobility is lowest when the Fermi wavelength is of the order of the correlation length ($k_{F} L \sim \pi/2$), but is lowered by Stark-effect scattering at {\em all} correlation lengths.  In Fig \ref{Fig2}(c), the mobility is plotted against the field $F_{w}$ for three different QW widths for fixed $n_{s}$ and interface roughness parameters.  It shows that wider QWs suffer more severely from Stark-effect scattering.  Thus, making QWs wider to reduce interface roughness scattering is not without penalties, especially if $F_{w}$ is large.

However, for very large fields and for wide QWs, the net scattering potential in Eq. \ref{scatpotential} may become of the order of intersubband energies ($E_{2} - E_{1} = 3 \pi^2 \hbar^2 /2m^{\star} L_{w}^2$) or other 1st-order energies of the `unperturbed' QW Hamiltonian in the scattering problem.  In such situations, it no longer suffices to treat the Stark-effect term as a perturbation; is more prudent to absorb the field $F$ directly into the unperturbed Hamiltonian.  To do so, we assume the unperturbed Hamiltonian of the form $H_{0} = -\hbar^{2}\partial^{2}/\partial z^{2} + e F z$, which yields eigenvalues $E_{n} = (\hbar^2/2m^{\star})^{1/3}(3\pi eF /2)^{2/3}(n+3/4)^{2/3}$ for the $n$th eigenstate, with corresponding eigenfunctions as Airy functions $\psi_{n}(z) = Ai[ (2 m^{\star} / \hbar^{2}) ( e F z  - E_{n}) ]$ \cite{AndoRMP82}.  As shown in \cite{AndoRMP82, BookDavis}, the Airy-eigenfunction can be closely approximated by the Fang-Howard variational function $\psi_{1}(z) = \sqrt{(b^{3}/2)}ze^{-bz/2}$ for the ground state, where $b=(33 m^{\star} e^2 n_s / 8 \hbar^2 \epsilon_{s})^{1/3}$ is the variational parameter of inverse length unit.  We note that whereas the Airy function assumes a constant electric field, the Fang-Howard wavefunction leads to a variable electric field which peaks at the heterojunction; this is indeed required by electrostatics.  This peak field is given simply by Gauss's law: $F_{\pi} = (e/\epsilon_{s}) \cdot \int_{0}^{\infty} n_{s} | \psi(z) |^{2} dz  = e n_{s} / \epsilon_{s}$.  The rough interface is located precisely where the field peaks, and the scattering potential is then given by $W(r) = - e F_{\pi} \Delta(r)$.

Using the 2DEG envelope function $\langle r | {\bf k} \rangle = [ \sqrt{(b^{3}/2)} z e^{-bz/2} ] e^{i {\bf k \cdot r}}$, the scattering matrix element $\langle {\bf k} | W(r) | {\bf k'} \rangle$ leads to $|W(q)|^2= (e F_{\pi})^{2} |\Delta(q)|^{2} $.  We note the similarity with the Stark-effect scattering result for the square quantum well derived earlier, but with $eF_{\pi}$ serving as the effective force $\mathcal{F}$.  Thus, the calculation of scattering rate and mobility for this case is done using the same expression as in Eq. \ref{sqwellresult}, but with $\mathcal{F} \rightarrow eF_{\pi}$, and a modified screening function.  The screening function is $\epsilon_{2D}(q) = 1+ G(q) \cdot q_{TF}/q $, where $G(q) = (2\eta^3+3\eta^2+3\eta)/8$  is the form factor with $\eta = b/(b + q)$ \cite{HirakawaPRB87}.  The IR-limited mobility thus degrades as the square of the peak field, $\mu_{IR} \sim 1/F_{\pi}^{2}$.  The total RT mobility $\mu= (\mu_{IR}^{-1}+ \mu_{POP}^{-1}+ \mu_{AP}^{-1})^{-1}$ of the 2DEGs can be calculated by a combination of polar optical phonon (POP) and acoustic phonon (AP) and IR scattering using Matthiessen's rule \cite{HirakawaPRB87}.  Very high polarization-induced fields exist in AlN/GaN polar heterostructures.  In such structures, the effect of increased IR scattering at room temperature reduces an intrinsic phonon-limited mobility of $\sim 2200 $ cm$^{2}$/V.s for $F_{\pi}$= 1.8 MV/cm ($n_{s} \sim 10^{13}/$ cm$^2$) to $ \sim 1680$ cm$^{2}$/V.s for $F_{\pi}$ =5.5 MV/cm ($n_{s} \sim 3 \times 10^{13}/$ cm$^2$).  These numbers are in good agreement with experiments \cite{Woodbook,Caopss08}, and indicate the strong mobility degradation by Stark-effect scattering.  In a HEMT-type device, the implication is that the electron mobility will initially increase as the gate pinches off the channel due to the reduction of IR/Stark scattering, but saturate below a certain density due to intrinsic phonon scattering limitations.

\par
In summary, this letter identifies and quantitatively evaluates the effect of Stark-effect scattering in the presence of an electric field on electron mobility in rough quantum wells.  When the field is small, a perturbative treatment shows that mobility reduces as $\mu_{IR} \sim L_{w}^{6}$ for thin wells ($L_{w} < L_{w}^{0}$), but switches over to $\mu_{IR} \sim L_{w}^{-6}$ above this critical width.  The implication is that Stark-effect scattering enforces a window of QW widths for high mobility.  On the other hand, for QWs where the field is too high to be treated perturbatively (such as in highly polar AlN/GaN QWs), the IR limited mobility degrades as the square of the peak electric field ($\mu_{IR} \sim 1/ F_{\pi}^{2}$) resulting in low mobilities for high carrier densities.  Since Stark-effect scattering feeds off interface roughness, it can be reduced by making smoother interfaces.  For a given interface roughness, it can be reduced by careful band-diagram engineering such that the field in the QW or at the heterojunction is minimized.

The authors acknowledge discussions with A. Konar, and financial support from NSF (ECCS) and the DARPA (NEXT) program.

%\bibliography{aipsamp}% Produces the bibliography via BibTeX.

\end{document}